%% file: main.tex
\documentclass{article}
\usepackage[preprint, nonatbib]{nips}

\usepackage{amsmath,amssymb,amsfonts}
\usepackage{algorithmic}

\usepackage{graphicx}
\usepackage{epstopdf}

\usepackage{textcomp}
\usepackage{xcolor}
\usepackage{mathtools}
\usepackage{comment}
\usepackage{caption}
\usepackage{subcaption}
\usepackage{tabularx,booktabs}
\usepackage{slashbox}
\usepackage{booktabs,caption}
\usepackage[flushleft]{threeparttable}
\usepackage{multirow}

\usepackage{amsmath}
\usepackage{color}
\usepackage{comment}

\usepackage{mathtools}

\usepackage{overpic}

\usepackage{tabularx} 

\usepackage{wrapfig}
\usepackage{wrapfig, blindtext}

\usepackage{cases}
\usepackage{empheq}

\usepackage{url}

\usepackage{tablefootnote}
\usepackage{footnote}

\newcolumntype{C}[1]{>{\centering\arraybackslash}p{#1}}

\author{
  Tian Guo 
  \thanks{
  Correspondence to \texttt{tig@ram-ai.com}
  }
  \,\,\,\,\,\,\,\,\,\,\,\,\,\,
  Emmanuel Hauptmann \\
  Systematic Equities Team 
  \\
  RAM Active Investments, Geneva, Switzerland 
}

\begin{document}

\title{
Fine-Tuning Large Language Models for Stock Return Prediction Using Newsflow 
}
\maketitle


\begin{abstract}
Large language models (LLMs) and their fine-tuning techniques have demonstrated superior performance in various language understanding and generation tasks.
This paper explores fine-tuning LLMs for stock return forecasting with financial newsflow.
In quantitative investing, return forecasting is fundamental for subsequent tasks like stock picking, portfolio optimization, etc. 
We formulate the model to include text representation and forecasting modules. 
We propose to compare the encoder-only and decoder-only LLMs, considering they generate text representations in distinct ways. 
The impact of these different representations on forecasting performance remains an open question.
Meanwhile, we compare two simple methods of integrating LLMs' token-level representations into the forecasting module.
The experiments on real news and investment universes reveal that:
(1) aggregated representations from LLMs' token-level embeddings generally produce return predictions that enhance the performance of long-only and long-short portfolios;
(2) in the relatively large investment universe, the decoder LLMs-based prediction model leads to stronger portfolios, whereas in the small universes, there are no consistent winners.
Among the three LLMs studied (DeBERTa, Mistral, Llama), Mistral performs more robustly across different universes;
(3) return predictions derived from LLMs' text representations are a strong signal for portfolio construction, outperforming conventional sentiment scores.
\end{abstract}

\input{introduction}

\input{related_work}

\input{model}

\input{experiment}
\input{conclusion}

\bibliographystyle{plain}
\bibliography{reference}

\clearpage
\appendix
\input{appendix}

\end{document}

%% file: introduction.tex
\section{Introduction}

Quantitative investing relies on extracting quantitative features or signals from various data sources including market prices, economic indicators, financial text, etc., to build and optimize investment portfolios~\cite{fama1996multifactor, ang2014asset}.
In recent years, the use of text data for quantitative investing has grown significantly, thanks to the advancement of natural language processing (NLP) techniques~\cite{xu2018stock,sawhney2020deep,qin2019you}.
In particular, large language models (LLMs) have demonstrated superior performance on various language understanding and generation tasks~\cite{he2021debertav3,behnamghader2024llm2vec, jiang2023mistral, touvron2023llama}, and the fine-tuning technique allows for adapting the pre-trained LLMs to fit investing-related applications~\cite{hu2021lora, ding2023parameter}.

This paper is focused on return forecasting with financial news for stock picking.
Return forecasting is useful for picking stocks with profit potentials to include in portfolios.
Financial news reports on events and announcements related to companies, industries, the economy, etc., and shows notable predictive power for stock future performance in previous studies~\cite{liu2018hierarchical,hu2018listening}.

The conventional way of applying financial news data to stock picking involves a multi-step extraction-and-validation process as illustrated in Fig.~\ref{fig:compare}(a), i.e., formulating the numerical features (e.g., sentiments, topics, popularity, etc.) with the expectation that these features have a predictive relationship with stock future performance (e.g., forward return, volatility, etc.)~\cite{allen2019daily,shapiro2022measuring}, developing the calculation processes or machine learning models to extract features from the news (e.g., train a financial sentiment classification model), and validating the predictive power of extracted features by statistical analysis or building forecasting models.
This process might be time-consuming and require additional data (e.g., labeled financial sentiment data) and continuous refinements.

LLMs generate numerical representations (or embeddings) of text that capture semantic relations, and these representations can naturally serve as features for forecasting tasks.
Given this intuition, this paper explores direct news-to-return prediction through fine-tuning LLMs.
Fig.~\ref{fig:compare} illustrates the difference between the conventional feature extraction-and-validation process and our LLM-based news-to-return process.
Though some previous works attempted to use text embedding for forecasting~\cite{liu2018hierarchical,wang2019ean,qin2019you, guo2020esg2risk}, few works have explored the potential of fine-tuning LLMs for stock return forecasting with newsflow. 
Moreover, this paper has the contribution as follows:
\begin{figure}[t]
  \begin{center}
    \includegraphics[width=0.8\textwidth]{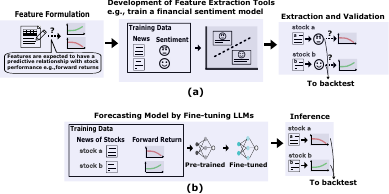}
  \end{center}
  \caption{
  Comparison of different workflows of utilizing financial news for stock picking.
  (a) Conventional feature extraction-and-validation process, e.g., financial sentiments.
  (b) News-to-return forecasting by fine-tuning LLMs.
  }
  \label{fig:compare}
\end{figure}

\begin{itemize}

\item We design an LLM-based return prediction model comprising the text representation and the forecasting modules.

\item We hypothesize that the text representations from encoder-only and decoder-only LLMs will perform differently due to their distinct methods of encoding text sequences during pre-training and fine-tuning; 
thus we propose to compare the encoder-only (DeBERTa) and decoder-only LLMs (Mistral, Llama3) as the representation module of the prediction model. 

\item Considering that LLM-generated text representations are at the token level, we present two simple methods to integrate token representations into the forecasting module: bottleneck representations and aggregated representations.

\item We perform experiments on real financial news and various investment universes.
In addition to evaluating prediction errors, we assess two types of portfolios built on return predictions through backtesting in out-of-sample periods.
The experimental comparison between encoder-only and decoder-only LLMs, as well as between bottleneck and aggregated representations, offers insights for identifying suitable text representations for different investing strategies and markets.

\end{itemize}

%% file: related_work.tex
\section{Related Work}

Numerous works have investigated using financial text data for forecasting tasks.
Previous works mostly used word-level embedding techniques lacking the ability of contextual modeling.
\cite{weng2018predicting, xu2018stock} extracted the sentiment score from financial newsflow, social media, and tweets for stock price predicting.
\cite{liu2018hierarchical, hu2018listening} explored learning numeric representations of financial news by attention mechanisms for modeling stock movements.
\cite{wang2019ean} studied combining sentiment and text representations for return prediction.
\cite{chen2019group} studied the embedding aggregate strategy of news for forex prediction.

The advent of LLMs and related techniques provides a new powerful way of using text data for forecasting tasks in quantitative investing~\cite{zhao2023survey, li2023large}.
LLMs can be broadly categorized into three main types.
Encoder-only models such as BERT (Bidirectional Encoder Representations from Transformers)~\cite{devlin2019bert} and DeBERTa (Decoding-enhanced BERT with disentangled attention)~\cite{he2020deberta, he2021debertav3}, focus on learning contextual embeddings for input text.
Decoder-only models like GPT-3 (Generative Pre-trained Transformer 3)~\cite{radfordimproving} and Mistral~\cite{jiang2023mistral} are trained to generate text by predicting the next token in a sequence.
Encoder-decoder models including T5 (Text-To-Text Transfer Transformer)~\cite{lewis2019bart} and BART (Bidirectional and Auto-Regressive Transformers)~\cite{raffel2020exploring} are a mix of both encoder and decoder architectures and suitable for sequence-to-sequence tasks such as machine translation, summarization, and question-answering.

LLMs are pre-trained on vast amounts of text data to learn general language patterns.
Following pre-training, there are two main approaches to applying LLMs to downstream tasks.
The prompt technique is to design specific inputs to guide the pre-trained LLM to produce the desired output without modifying the LLM's parameters~\cite{radford2019language, brown2020language, kojima2022large}.
The second approach is to fine-tune LLMs by adjusting the pre-trained LLM's parameters to adapt to specific tasks~\cite{gunel2020supervised, wei2021finetuned, ding2023parameter, chung2024scaling}.
In particular, parameter-efficient fine-tuning techniques have gained popularity~\cite{hu2021lora, ding2023parameter, liu2024dora}.
For instance, LoRA (Low-Rank Adaptation)~\cite{hu2021lora} introduces low-rank adaptations to the pre-trained model parameters, thereby reducing
the computational and memory overhead of fine-tuning.

Some recent works use LLMs as feature extractors to obtain predictive signals from text.
Authors in \cite{araci2019finbert, liu2021finbert} explored the fine-tuning of pre-trained LLMs to provide more accurate financial sentiment analysis.
Instead of fine-tuning LLMs, \cite{wang2024llmfactor} extracted factors from the financial news and price history by prompts on generative LLMs.
\cite{kim2024financial} used chain-of-thought prompts~\cite{wei2022chain} on generative LLMs to analyze financial statements.

Unlike existing works that extract features from text using LLMs, this paper focuses on fine-tuning LLMs to directly model the relation between financial text and stocks' future performance, i.e., newsflow and forward return.
Meanwhile, we evaluate the text representations from different types of LLMs to study their different effectiveness for the return forecasting task.

%% file: model.tex
\section{From Financial Newsflow to Stock Portfolios through LLMs}

\subsection{Problem Statement}\label{sec:problem}

Assume an investment universe consisting of a set of stocks denoted by $\mathcal{U} = \{ s \}_{s=1}^S $, where $s$ represents the stock index.
In quantitative investing, the stock-picking process selects a subset of the universe as the investing portfolio based on quantitative criteria. 
As market conditions and various information change, the stock-picking process is repeatedly performed to update or rebalance the portfolios at (regular) time intervals, e.g., weekly, monthly, etc.

\begin{figure}[!htbp]
  \begin{center}
    \includegraphics[width=0.75\textwidth]{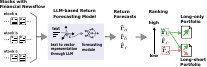}
  \end{center}
  \caption{
    Illustration of the LLM-based return forecasting model for the stock-picking process.
    Assume an investment universe of 3 stocks denoted by $a, b, c$.
    Each stock has an associated list of news.
    Then, given the return forecasts and ranks, stocks can be selected into long-only or long-short portfolios.
  }
  \label{fig:portfolio}
\end{figure}
This paper is interested in predicting stock returns with news for stock picking.
Specifically, let $r_{s,t + n} \in \mathbb{R}$ be the $n$-step forward return of stock $s$ w.r.t. timestep $t$. 
The textual content of financial news reported at time $i$ and w.r.t. stock $s$ is denoted by $\mathbf{x}_{s, i}$, a list of text tokens.
At time $t$, the news text available for predicting $r_{s,t + \ell}$ in a look-back time window $W$ is $\{ \mathbf{x}_{s, i} \}_{ i \in \mathcal{T}_{s, <t} }$  where $\mathcal{T}_{s, <t}$ represents the set of timesteps of available news.

Considering the large sequence length that LLMs can process nowadays~\cite{zhao2023survey, li2023large}, we concatenate the set of news in the look-back window into one sequence denoted by $\mathbf{X}_{s, <t} =  \oplus \{ \mathbf{x}_{s, i} \}_{ i \in \mathcal{T}_{s, <t} }$, where $\oplus$ denotes the concatenation operation.
Next, we formulate the return forecasting model as a composite structure of a text representation module and a forecasting module as defined in Eq.~\ref{eq:model}:
\begin{align}\label{eq:model}
  \hat{r}_{s, t + \ell} = f \circ g\left(\mathbf{X}_{s,<t}\right)
\end{align}

We aim to explore realizing Eq.~\ref{eq:model} by jointly fine-tuning a pre-trained LLM as $g(\cdot)$ and training a dense layer as $f(\cdot)$.
In particular, Eq.~\ref{eq:model} is a sequence-level task requiring the text representation module $g \colon \mathbf{X}_{s,<t} \mapsto \mathbf{h}_{s, <t} $ to encode the sequence $\mathbf{X}_{s, <t}$ into a numerical vector $\mathbf{h}_{s, <t} \in \mathbb{R}^{D}$.
Then, the forecasting module $f \colon \mathbf{h}_{s, <t} \mapsto \hat{r}_{s,t} $ transforms $\mathbf{h}_{s, <t}$ to the return forecast.
We train the model using a set of data instances pooled from individual stocks and associated news, i.e., $\{ (r_{s,t + \ell}, \mathbf{X}_{s,<t}) \}_{s \in \mathcal{U}, t \in \mathcal{T} }$ where $\mathcal{T}$ represents the timestamps in the training period.

At test time, besides evaluating prediction errors such as the root mean square error (RMSE), we implement the return prediction-based stock picking to construct long-only and long-short portfolios which are subsequently backtested.
This process is illustrated in Fig.~\ref{fig:portfolio}.

\textit{Long-Only Portfolios} is intended to include stocks with the expectation of a price rise above the universe average.
In practice, it is built by ranking the stocks based on the return forecasts and selecting the top-K stocks.
$K$ is usually chosen according to the decile or quantile of the universe, e.g., $10\%$ of the total number of stocks.

\textit{Long-Short Portfolios} includes both the stocks with the expectation of a price rise and drop.
For the stocks with a price drop expectation, the portfolio can profit by selling them at the present price and repurchasing them at a lower price in the future.
In this paper, the long-short portfolio is built by including the top-K and bottom-K stocks based on the forecast ranks.

\subsection{
Methodology
}

Transformer-based LLMs can be categorized into three main types: encoder-only, decoder-only, and the hybrid encoder-decoder. 
All these LLMs transform text into high-dimensional vector representations, however, their different pre-training objectives lead to text representations with varying implications.

In the following, we describe the text representation difference in encoder-only and decoder-only LLMs.
Then, we present two simple methods of integrating the token-level representations from LLMs into the forecasting module.
These methods introduce no additional parameters to learn and provide a clear comparison of the native representations of different LLMs for return forecasting.

\textbf{Encoder-only LLMs vs. Decoder-only LLMs.}
Given a sequence of text tokens $\mathbf{X} = \{ x_1, \cdots, x_L \}$, LLMs output a sequence of vector representations $\{ \mathbf{h}_1, \cdots, \mathbf{h}_L \}$ corresponding to the input tokens.
However, as presented below, the vector representations from encoder-only and decoder-only LLMs encode the different parts of the input sequence.

Pre-training an encoder LLM is mostly based on masked-language modeling~\cite{devlin2019bert, lan2019albert, he2020deberta}.
Concretely, it prepares a training text sequence $\mathbf{X}$ by randomly masking some tokens, leading to $\mathbf{\hat{X}} = \{ x_\text{mask} \,\, \text{if}\, i \in \mathcal{M}  \,\,\text{else}\,\,  x_i \,\, \forall \, i = 1, \cdots, L \}$.
$\mathcal{M} \subset \{1, \cdots, L\}$ represents the indices of tokens to mask.
The mask token $x_{\text{mask}}$ is a special token without concrete meaning and plays as the placeholder.
Then, the pre-training objective is to predict the masked tokens, i.e., maximizing the likelihood of masked tokens as:
\begin{align}
  \log p\big( \{x_m\}_{m \in \mathcal{M}} \,|\, \mathbf{\hat{X}} \big) 
  = \sum_{m \in \mathcal{M}} \log p( x_m \,|\, \mathbf{X}_{<m}, x_\text{mask}, \mathbf{X}_{>m} ) 
  \approx \sum_{m \in \mathcal{M}} \log p( x_m \,|\, \mathbf{h}_{m} )
  \label{eq:encoder}
\end{align}

In Eq.~\ref{eq:encoder}, $\mathbf{X}_{<m} = \{x_1, \cdots, x_{m-1}\} $ and $\mathbf{X}_{>m}  = \{ x_m, \cdots, x_{L}\}$ represent the tokens before and after $x_{m}$.
Maximizing Eq.~\ref{eq:encoder} encourages the representation $\mathbf{h}_{m}$ to incorporate both the left and right contexts, i.e., $\mathbf{X}_{>m}$ and $\mathbf{X}_{<m}$, for predicting the masked token.
Particularly, in the attention mechanism of Transformers, $\mathbf{h}_{m}$ is derived based on the similarities between the mask token $x_{\text{mask}}$ and the context tokens $\mathbf{X}_{>m}$ and $\mathbf{X}_{<m}$.

On the other hand, a decoder-only LLM models an input sequence autoregressively using the next-token prediction task~\cite{radfordimproving, touvron2023llama}.
The pre-training objective function is defined in Eq.~\ref{eq:decoder}:
\begin{align}
  \log p( x_1, \cdots, x_L | \mathbf{\check{X}}) 
  = \sum_{i = 1, \cdots, L} \log p( x_i \,|\, \mathbf{X}_{ < i} ) 
  \approx  \sum_{i} \log p( x_i \,|\, \mathbf{h}_{i-1} ) \label{eq:decoder}
\end{align}

For modeling the first token, the practical way is to add a Beginning-of-Sequence (BOS) token, i.e., $\mathbf{\check{X}} =  x_{\text{bos}} \oplus \mathbf{X} $.
Similar to the mask token, the BOS token has no concrete meaning.
The representation $\mathbf{h}_{i-1}$ encodes the information from already seen tokens and is derived based on the relation between $x_{i-1}$ and $\mathbf{X}_{< i-1} = \{ x_1, \cdots, x_{i-2}\}$.

\textbf{Bottleneck Representations vs. Aggregated Representations.}
As LLMs output the token-level vector representations, to obtain a representation encoding the sequence, the idea of bottleneck representation is to push LLMs to compress the sequence information into a single vector representation during fine-tuning~\cite{yang2019leveraging, wang2023simlm, wang2023improving}.

In practice, this is achieved by appending an End-of-Sequence (EOS) ${x}_{\text{EOS}}$ to the input sequence, e.g., $\mathbf{X}_{s, <t} \oplus {x}_{\text{EOS}}$.
As ${x}_{\text{EOS}}$ is constant across sequences, its vector representation $\mathbf{h}_{\text{EOS}}$ depends on the real tokens of the sequence.
During fine-tuning, $\mathbf{h}_{\text{EOS}}$ is fed into the forecasting module as shown in Eq.~\ref{eq:bottleneck}.
The backpropagation process propels $\mathbf{h}_{\text{EOS}}$ to summarize real tokens's representations through the forecasting module.
\begin{align} \label{eq:bottleneck}
    \hat{r}_{s, t + \ell} = f(\mathbf{h}_{\text{EOS}})
\end{align}
The bottleneck representation has different implications for encoder-only and decoder-only LLMs.
In encoder-only LLMs, the vector used for predicting is obtained based on the mask token and the real context tokens during the pre-training, as explained in Eq.~\ref{eq:encoder}.
As a result, appending an EOS token (identical to the mask token used in pre-training) aligns the fine-tuning with the pre-training.
This consistency might facilitate the EOS token representation to summarize sequence-level features effectively.
In decoder-only LLMs, the vector representation of each token is conditioned on the already-seen tokens; thus, the last token of a sequence naturally summarizes the whole sequence, making an additional EOS token redundant.

In experiments, we observed that appending the EOS token is more helpful for encoder-only LLMs. 
For a comparison on the same ground, we append EOS tokens for both encoder-only and decoder-only LLMs and leave the study on the different impacts of appending tokens to future work.

Meanwhile, considering the recent works on the representation collapse issue of the last token in certain conditions~\cite{barbero2024transformers}, we present a simple alternative to bottleneck representation, i.e., allowing the forecasting module to aggregate the representations of all tokens. 
This can be done using various methods like averaging, or sophisticated ones like attention mechanisms~\cite{lee2024nv}.
In this paper, we choose the simple averaging method, since it introduces no additional parameters to train and enables a clear comparison with the bottleneck representation.
\begin{align}
   \hat{r}_{s, t + \ell} = f\left( \frac{1}{L} \sum_{l} \mathbf{h}_{l} \right)
\end{align}
For encoder-only LLMs, the pre-training and fine-tuning discrepancy arises when using aggregated representations, because each token's representation is based on context and itself, instead of the mask token in pre-training.
For decoder-only LLMs, averaging all representations might lead to bias towards the early tokens of the input sequence.
This is because, in the autoregressive setting, the early tokens are repeatedly incorporated into the representations of all subsequent ones.

\textbf{Implementations.}
The text representation module and the forecasting module are respectively initialized by a pre-trained LLM and a dense layer.
Then, the training process jointly fine-tunes the LLM and learns the forecasting module to minimize the mean squared error (MSE) between the forecasts and true values.
We applied Low-Rank Adaptation (LoRA) to fine-tune LLMs~\cite{hu2021lora}.
Other techniques including gradient checkpointing, mixed precision training, and DeepSpeed are used to reduce GPU memory~\cite{rasley2020deepspeed}.

We experiment with one encoder-only LLM, i.e., DeBERTa~\cite{he2021debertav3}, and two different decoder-only LLMs, i.e., Mistral-7B and Llama3-8B base models~\cite{touvron2023llama,jiang2023mistral}.
DeBERTa is a recent encoder-only LLM that improves upon the BERT model with disentangled content and position embeddings.
Mistral-7B is a 7-billion-parameter decoder-only LLM that uses grouped query and sliding window attention to improve performance.
Llama3-8B is an 8-billion-parameter decoder-only LLM pre-trained on data mixed from different sources, e.g., multilingual, codes, etc., to improve the generalization ability.

%% file: experiment.tex
\section{Experiments}\label{sec:exp}

\textbf{Data.}
We use company-level financial newsflow data from 2003 to 2019 provided by a financial data vendor.
Each piece of news has an attribute including the company identifier(s) the news is primarily about.
Meanwhile, we have two investment universe datasets of the North American (NA), European (EU), and Emerging (EM) markets, which consist of dates, stock identifiers, and the true monthly forward returns of corresponding stocks and dates.
The training and validation data is from 2003 to 2014 for each universe, while the rest is for the out-of-sample testing data.
Each instance is built by linking an entry in the universe data to related news through the stock identifier and a look-back time window (e.g., one week).
Table~\ref{tab:data_stats} shows the data stats.

\textbf{Setup.}
We train the model only once and then apply the model to obtain the return predictions in the testing period.
We conduct the model training using a batch size of $32$, a learning rate of $1e$-$5$, and a warmup phase of $100$ steps followed by a linear decay.
To fine-tune LLMs, we applied Low-Rank Adaptation (LoRA) with rank $4$ to all linear layers.
We employ a maximum context length of $4$k for all LLMs used in experiments. 
All models are trained for $10$ epochs on $2$ A$100$ GPUs.

The long-only portfolio is built by taking the stocks with the return predictions falling in the top ($9$th) decile of prediction rankings.
The long-short portfolios take the stocks in the top ($9$th) and bottom ($0$th) deciles.
The stocks in all portfolios are equally weighted.

We perform backtesting to evaluate the portfolios in monthly rebalancing.
It stimulates the trading of monthly constructed portfolios and reports the cumulative return chart and performance statistics like annualized returns and Sharpe ratios in the testing period.
When backtesting the long-only and long-short portfolios, besides comparing the portfolios built on return predictions by different LLMs, we also compare them with the sentiment-based portfolio construction.
Specifically, FinBERT is a fine-tuned BERT (Bidirectional Encoder Representations from Transformers) for financial sentiment analysis~\cite{araci2019finbert}.
FinVader is a dictionary-based method with a financial sentiment lexicon~\cite{hutto2014vader, FinVADER}.
The sentiment-based portfolios are built using the same method but with sentiment values as the ranking criteria.

\textbf{Metrics.}
As mentioned in the problem statement of Sec.~\ref{sec:problem}, the downstream stock picking for building portfolios is based on the deciles of forecasts; thus we report three decile-wise metrics to align with downstream scenarios, i.e., decile RMSE, decile precision, and decile return.
The decile return is the actual return of stocks allocated to the decile based on predictions and is directly related to the portfolio performance.
Analyzing the decile return along with the decile RMSE and precision provides insights into the relation between portfolio performance and prediction accuracy.

Specifically, at each date in the testing data, we group the predictions with the true returns into deciles based on the ranking of predictions (i.e., the highest predictions are in the top $9$th decile and the lowest ones are in the bottom 0th decile). 
Then, with the true and predicted returns in each decile across dates, we calculate the decile RMSE, decile precision, and decile return.
The decile precision is the percentage of the true returns whose decile based on the ranking of true values is equal to the current decile.
It is related to the portfolio performance, because, for instance, a high precision of the top decile implies that a high proportion of stocks in this decile has a high true forward return, thereby benefiting the portfolio including stocks from the top decile.

For portfolio backtesting, we report the cumulative return charts and performance statistics like annualized returns and Sharpe ratios in the testing period.

\begin{figure}[!htbp]
  \begin{center}
    \includegraphics[width=0.85\textwidth]{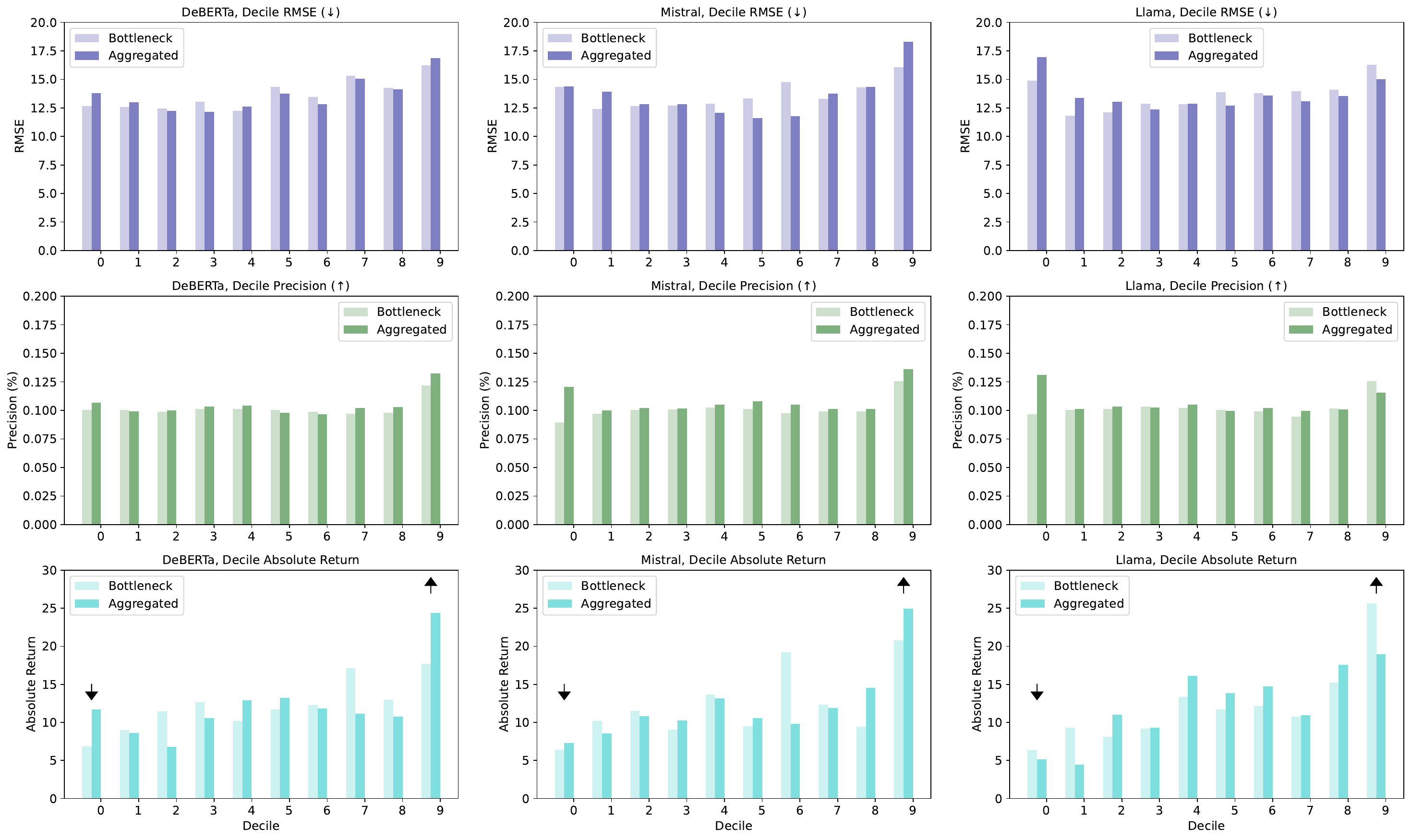}
  \end{center}
  \caption{
    Decile Performance of Bottleneck and Aggregated Representations in the North American Universe (best viewed in color).
    Top Row: Decile RMSE.
    Middle Row: Decile Precision.
    Bottom Row: Decile Return.
    The up (or down) arrow indicates the higher (or lower) values are desirable.
  }
  \label{fig:decile_na}
\end{figure}

\begin{table*}[!htbp]
  \centering
  \caption{
  Statistics of Portfolios in the North American Universe. 
  The Universe Equally-Weighted represents the universe performance reported under the Long-only Portfolio column.
  }
  \resizebox{1.\textwidth}{!}{
  \begin{tabular}{|c|c|c|c|c|}
    \hline
    \multirow{2}{*}{} & \multicolumn{2}{|c|}{Long-only Portfolio} & \multicolumn{2}{|c|}{Long-short Portfolio} \\
    \cline{2-5}
    & Ann. Return \% ($\uparrow$) & Sharpe Ratio ($\uparrow$) & Ann. Return \% ($\uparrow$) & Sharpe Ratio ($\uparrow$) \\
    \hline
    Universe Equally-Weighted      & 9.76  & 0.68   & $-$  & $-$ \\
    Sentiment\_FinVader            & 12.26 & 0.72   & 2.92 & 0.39\\
    Sentiment\_FinBert             & 20.64 & 1.22   & 8.81 & 0.92 \\
    \hline
    DeBERTa\_Bottleneck  & 17.47 & 0.96  & 10.83  & 0.94 \\
    DeBERTa\_Aggregated  & 25.15 & 1.20  & 12.87  & 1.07 \\
    Mistral\_Bottleneck  & 21.27 & 1.15  & 15.08  & 1.49 \\
    Mistral\_Aggregated  & 25.38 & 1.12  & 18.30  & 1.26 \\
    Llama\_Bottleneck    & 27.00 & 1.32  & 20.46  & 1.49 \\
    Llama\_Aggregated    & 18.86 & 1.00  & 14.29  & 1.30 \\
    \hline
  \end{tabular}
  }
  \label{tab:portfolio_na}
\end{table*}

\textbf{Results.} In the following, we present and discuss mainly the results of the NA universe.
The results of the EU and EM universe are in the Appendix section.

\begin{figure}[htbp]
  \begin{center}
    \includegraphics[width=0.85\textwidth]{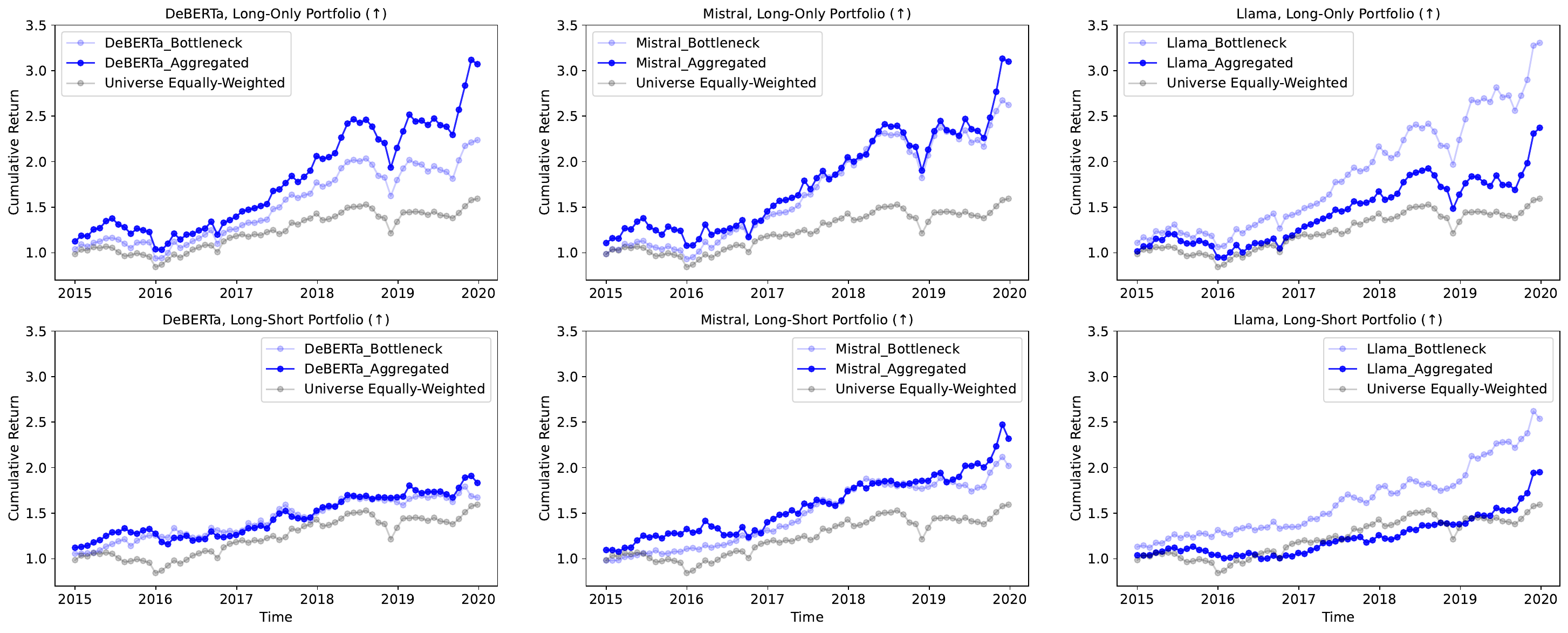}
  \end{center}
  \caption{
  Cumulative Return Charts of the Portfolios based on Bottleneck and Aggregated Representation Models in the North American Universe (best viewed in color).
  Top Row: Long-only Portfolios. 
  Bottom Row: Long-short Portfolios. 
  }
  \label{fig:portfolio_na}
\end{figure}

\underline{Bottleneck Representations vs. Aggregated Representations:} 
In Fig.~\ref{fig:decile_na}, we compare the bottleneck and aggregated representations for the three LLMs in the North American universes through the decile RMSE, precision, and returns.
Each column of Fig.~\ref{fig:decile_na} corresponds to a LLM.
Meanwhile, Fig.~\ref{fig:portfolio_na} shows the cumulative return charts of portfolios and Table~\ref{tab:portfolio_na} reports the detailed performance stats of portfolios.

In the bottom row of Fig.~\ref{fig:decile_na}, the returns from the 0th decile to the 9th decile generally present an upward trend, implying that overall the return predictions are aligned with actual future performance.
Moreover, we are particularly interested in the top $9$th and bottom $0$th deciles as they are the main constituents of portfolios.
For the top $9$th decile, the aggregated representation model generates a higher return and benefits the long portfolio, except for Llama.
For the EU and EM universe, as presented in the Appendix section, the aggregated representation model consistently outperforms the bottleneck one.

Interestingly, the higher returns do not necessarily imply low RMSE in the $9$th decile.
For instance, in Fig.~\ref{fig:decile_na}, the aggregated representation model has a higher decile return, but a higher RMSE, in the $9$th decile corresponding to the long-only portfolio for DeBERTa and Mistral.
An explanation is that the $9$th decile is regarding predicting high-value returns and less accurate predictions of these returns might have high RMSE.
But, if the return prediction still falls into the $9$th decile as the true return, the corresponding decile return is retained.
In this case, the decile precision is more indicative of the decile return, for instance, in Fig.~\ref{fig:decile_na} the outperforming representations mostly have a higher precision in the $9$th decile.

As for the bottom $0$th decile, a lower return is preferred as the short side of a long-short portfolio benefits from stocks with underperforming forward returns.
In Fig.~\ref{fig:decile_na}, the aggregated representation model falls short of lowering the 0th decile's return for DeBERta and Mistral, however, Table~\ref{tab:portfolio_na} shows that the return and Sharpe ratios of long-short portfolios are mostly improved with aggregated representations compared to the bottleneck representations.

Meanwhile, in the 0th decile, there are complexities in how prediction errors translate to actual returns. 
For instance, for DeBERTa, the aggregated representation has higher RMSE and precision in the bottom $0$th decile, implying that some stocks with higher true returns are misallocated to the 0th decile by the prediction. 
As a result, the 0th decile return of the aggregated representation is higher.
However, when the aggregated representation of Llama has the same pattern in the bottom decile, the return is as low as expected. 
This might be because the high precision offsets the impact of misallocated high returns.

Fig.~\ref{fig:portfolio_na} visualizes the cumulative return of the portfolios using the bottleneck and aggregated representation models.
The performance of long-only and long-short portfolios correspond to the top and bottom deciles in Fig.~\ref{fig:decile_na}.
The return curves of the aggregated representation model are notably higher except for Llama.
As shown in the Appendix, the aggregated representation constantly outperforms the bottleneck representation for the EU and EM universes.

\begin{figure}[!htbp]
  \begin{center}
    \includegraphics[width=0.85\textwidth]{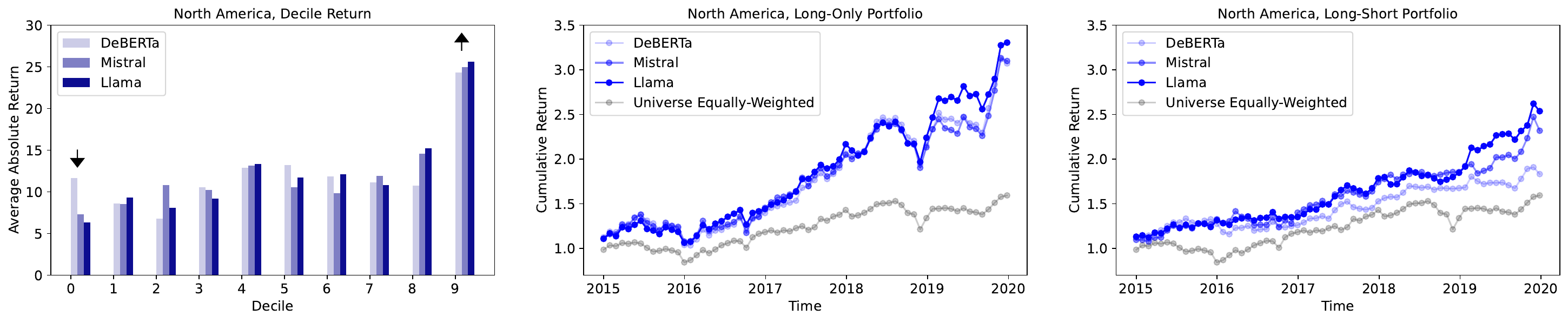}
  \end{center}
  \caption{
   Comparison of Encoder-only and Decoder-only LLMs with the Suited Representations in the North American Universe (best viewed in color).
  }
  \label{fig:compare_enc_dec_na}
\end{figure}

\underline{Encoder-only LLMs vs. Decoder-only LLMs:} Fig.~\ref{fig:compare_enc_dec_na} shows the comparison of encoder-only and decoder-only LLMs with the suitable representations for the NA universe, i.e., the aggregated representation for DeBERTa and Mistral, and the bottleneck representation for Llama. 
For the EU and EM universes in the Appendix, the aggregated representation is favored for all three LLMs.

The decile return in Fig.~\ref{fig:compare_enc_dec_na} exhibits that decoder-only Mistral and LLama generate high returns in the top 9th decile and lower returns in the bottom 0th decile, thereby leading to the outperforming long-only and long-short portfolios as shown in the cumulative return charts.
In particular, the performances of long-only portfolios are comparable among encoder and decoder LLMs, however, in long-short portfolios, the short side drags down the performance of the long side, especially for the encoder-only DeBERTa.
This highlights the importance of effective stock selection on both sides of the portfolio.
Meanwhile, all the prediction-based portfolios yield higher returns than the universe average.

\begin{figure}[!htbp]
  \begin{center}
    \includegraphics[width=0.85\textwidth]{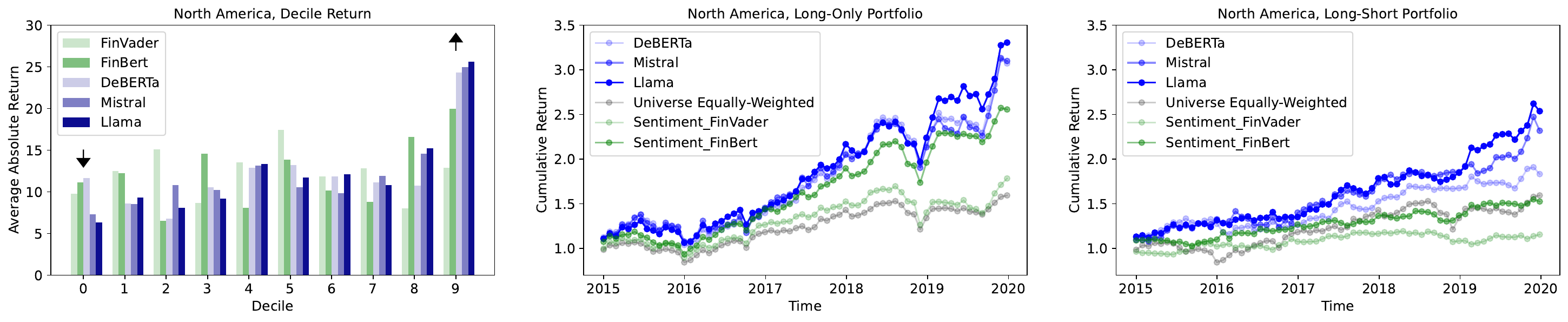}
  \end{center}
  \caption{
   Comparison with Sentiment-based Portfolios in the North American Universe (best viewed in color).
  }
  \label{fig:compare_senti_na}
\end{figure}

\underline{Prediction-based vs. Sentiment-based Portfolios:}
In this part, we compare the prediction-based portfolios with conventional sentiment-based portfolios.
Fig.~\ref{fig:compare_senti_na} shows the decile returns and the return charts of portfolios, and the performance statistics are in Table~\ref{tab:portfolio_na}.
The prediction-based portfolios are from the forecasting model with the suited representations, as in the above comparison of encoder-only and decoder-only LLMs.

In Table~\ref{tab:portfolio_na}, the prediction-based long-only and long-short portfolios outperform the sentiment-based portfolios both in returns and Sharp ratios.
In Fig.~\ref{fig:compare_senti_na}, the return charts of prediction-based portfolios are above the sentiment-based portfolios.
In particular, for the long-short portfolios, as shown in the return chart, the short side of the sentiment-based method negatively offsets the long side, leading to underperformance compared with the universe. 
In contrast, the prediction-based long-short portfolios have smoother return curves than the long-only portfolios, because the short side mitigates the overall portfolio's volatility. 
The outperformance of prediction-based portfolios suggests that the return prediction models capture more relevant information from text representations for future stock performance, leading to effective stock picking.

%% file: conclusion.tex
\section{Conclusion}

This paper focuses on return forecasting with financial newsflow for quantitative portfolio construction.
Unlike the conventional feature extraction-and-validation workflow, this paper explores fine-tuning LLMs to directly model the relationship between text representations and stock forward return.
Considering that different LLMs generate token-level representations in distinct ways, we compare the design choices on two aspects: the encoder-only versus decoder-only LLMs, and the bottleneck versus aggregated representations.

Our experiments are conducted on real financial news, various investment universes, and different portfolios.
The results reveal the key findings:
(1) aggregated representations from LLMs' token-level embeddings generally produce the return predictions that enhance the performance of long-only and long-short portfolios;
(2) in the relatively large investment universe, the decoder LLMs-based prediction model leads to stronger portfolios, whereas in the small universes, there are no consistent winners.
Among the three LLMs studied (DeBERTa, Mistral, Llama), Mistral performs more robustly across different universes;
(3) return predictions derived from LLMs' text representations are a strong signal for portfolio construction, outperforming conventional sentiment scores.

Several open questions remain for future research. 
For instance, it is unclear whether the underperformance of encoder-only DeBERTa in the large investment universe is due to the model size or other factors, and why DeBERTa has varying performance in different small universes.
Evaluating recently proposed large encoder-only LLMs \cite{wang2023improving,behnamghader2024llm2vec} would be an interesting follow-up.
Additionally, within the decoder-only LLM family, compared with Mistral's robust performance across investment universes, the reasons behind Llama's performance variation need further exploration.


%% file: appendix.tex
\section{Appendix}

\begin{table*}[htbp]
  \centering
  \caption{
  Statistics of Datasets. 
  }
  \resizebox{1.\textwidth}{!}{
  \begin{tabular}{|c|c|c|c|c|c|}
    \hline
    Universe       & \# of Stocks & Average \# of News per Instance  & \# of Training Instances & \# of Validating Instances & \# of Testing Instances \\
    \hline
    North America     & 630   & 2.5   & 366011 & 10167 & 241367 \\
    Europe            & 350   & 1.9   & 100403 & 10041 & 121705 \\
    Emerging Markets  & 370   & 2.6   &  71610 & 10231 & 183608 \\
    \hline
\end{tabular}
}
\label{tab:data_stats}
\end{table*}

\subsection{Results of the European Universe}

\begin{figure}[htbp]
    \centering
    \begin{subfigure}[t]{.85\textwidth}
        \centering
        \includegraphics[width=1.\textwidth]{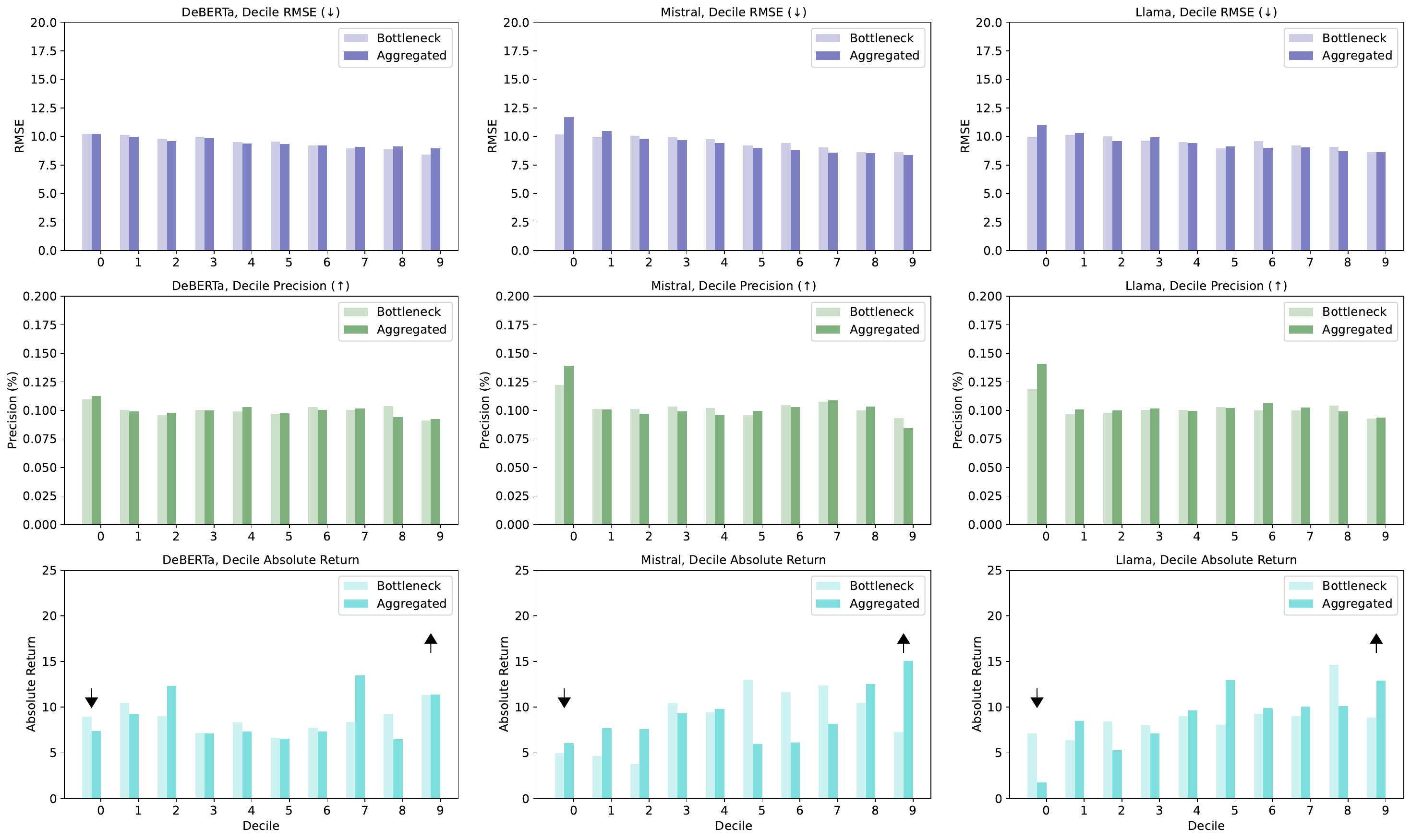}
    \end{subfigure}%
    \caption{
    Decile Performance of Bottleneck and Aggregated Representations in the European Universe (best viewed in color).
    Top Row: Decile RMSE.
    Middle Row: Decile Precision.
    Bottom Row: Decile Return.
    The up (or down) arrow indicates the higher (or lower) values are desirable.
    }
    \label{fig:decile_eu}
\end{figure}

\begin{table*}[htbp]
  \centering
  \caption{
  Statistics of Portfolios in the European Universe. 
  The Universe Equally-Weighted represents the universe performance reported under the Long-only Portfolio column.
  }
  \resizebox{1.\textwidth}{!}{
  \begin{tabular}{|c|c|c|c|c|}
    \hline
    \multirow{2}{*}{} & \multicolumn{2}{|c|}{Long-only Portfolio} & \multicolumn{2}{|c|}{Long-short Portfolio} \\
    \cline{2-5}
    & Ann. Return \% ($\uparrow$) & Sharpe Ratio ($\uparrow$) & Ann. Return \% ($\uparrow$) & Sharpe Ratio ($\uparrow$) \\
    \hline
    Universe Equally-Weighted  & 9.75  & 0.74 & $-$   & $-$ \\
    Sentiment\_FinVader            & 10.25 & 0.70 & 3.40  & 0.45 \\
    Sentiment\_FinBert             &  8.17 & 0.57 & -0.36 & 0.00 \\
    \hline
    DeBERTa\_Bottleneck  & 11.04 & 0.81 & 2.11 & 0.31 \\
    DeBERTa\_Aggregated  & 11.11 & 0.81 & 3.84 & 0.52 \\
    Mistral\_Bottleneck  &  6.40 & 0.48 & 1.94  & 0.26 \\
    Mistral\_Aggregated  & 15.12 & 1.02 & 9.07  & 1.04 \\
    Llama\_Bottleneck    &  8.20 & 0.62 & 1.25  & 0.17 \\
    Llama\_Aggregated    & 12.76 & 0.90 & 11.47 & 1.27  \\
    \hline
  \end{tabular}
  }
  \label{tab:portfolio_eu}
\end{table*}

\clearpage
\begin{figure}[ht!]
  \begin{center}
    \includegraphics[width=0.85\textwidth]{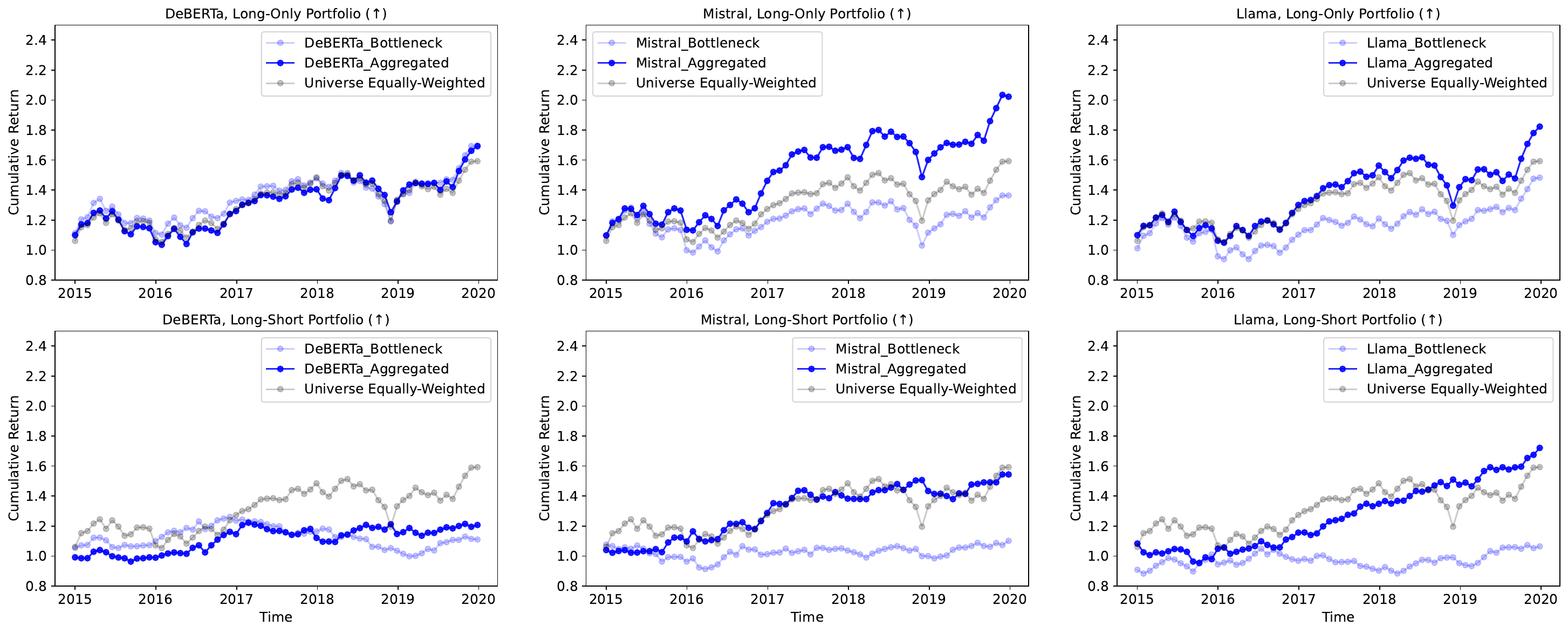}
  \end{center}
  \caption{
  Cumulative Return Charts of the Portfolios based on Bottleneck and Aggregated Representation Models in the European Universe (best viewed in color).
  Top Row: Long-only Portfolios. 
  Bottom Row: Long-short Portfolios. 
  }
  \label{fig:portfolio_eu}
\end{figure}

\begin{figure}[htbp]
  \begin{center}
    \includegraphics[width=0.85\textwidth]{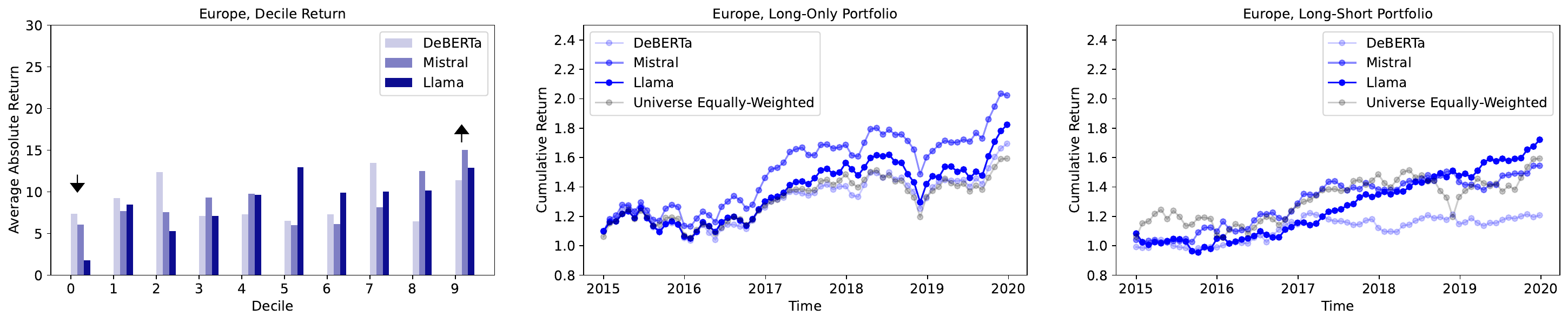}
  \end{center}
  \caption{
  Comparison of Encoder-only and Decoder-only LLMs with the Suited Representations in the European Universe (best viewed in color).
  }
  \label{fig:compare_enc_dec_eu}
\end{figure}

\begin{figure}[htbp]
  \begin{center}
    \includegraphics[width=0.85\textwidth]{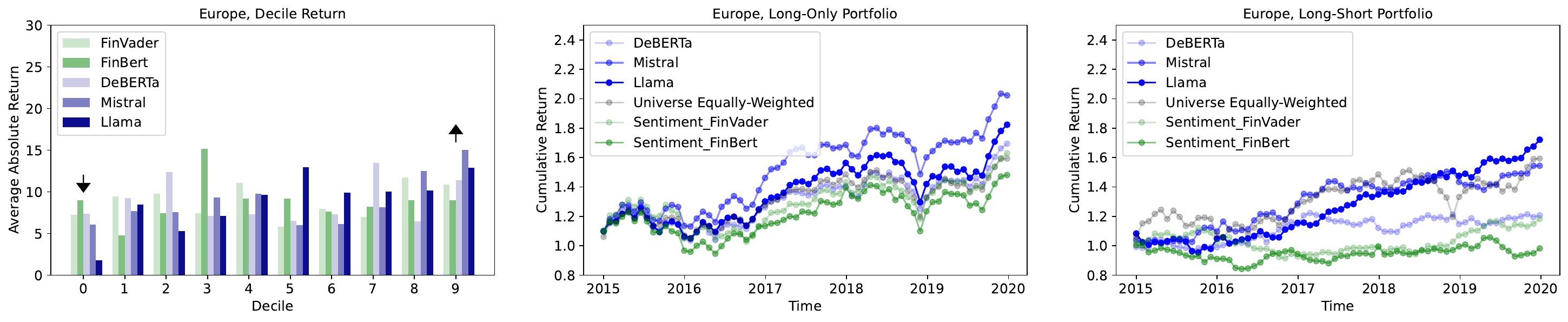}
  \end{center}
  \caption{
  Comparison with Sentiment-based Portfolios in the European Universe (best viewed in color).
  }
  \label{fig:compare_senti_eu}
\end{figure}

\clearpage
\subsection{Results of the Emerging Markets Universe}

\begin{figure}[htbp]
    \centering
    \begin{subfigure}[t]{.85\textwidth}
        \centering
        \includegraphics[width=1.\textwidth]{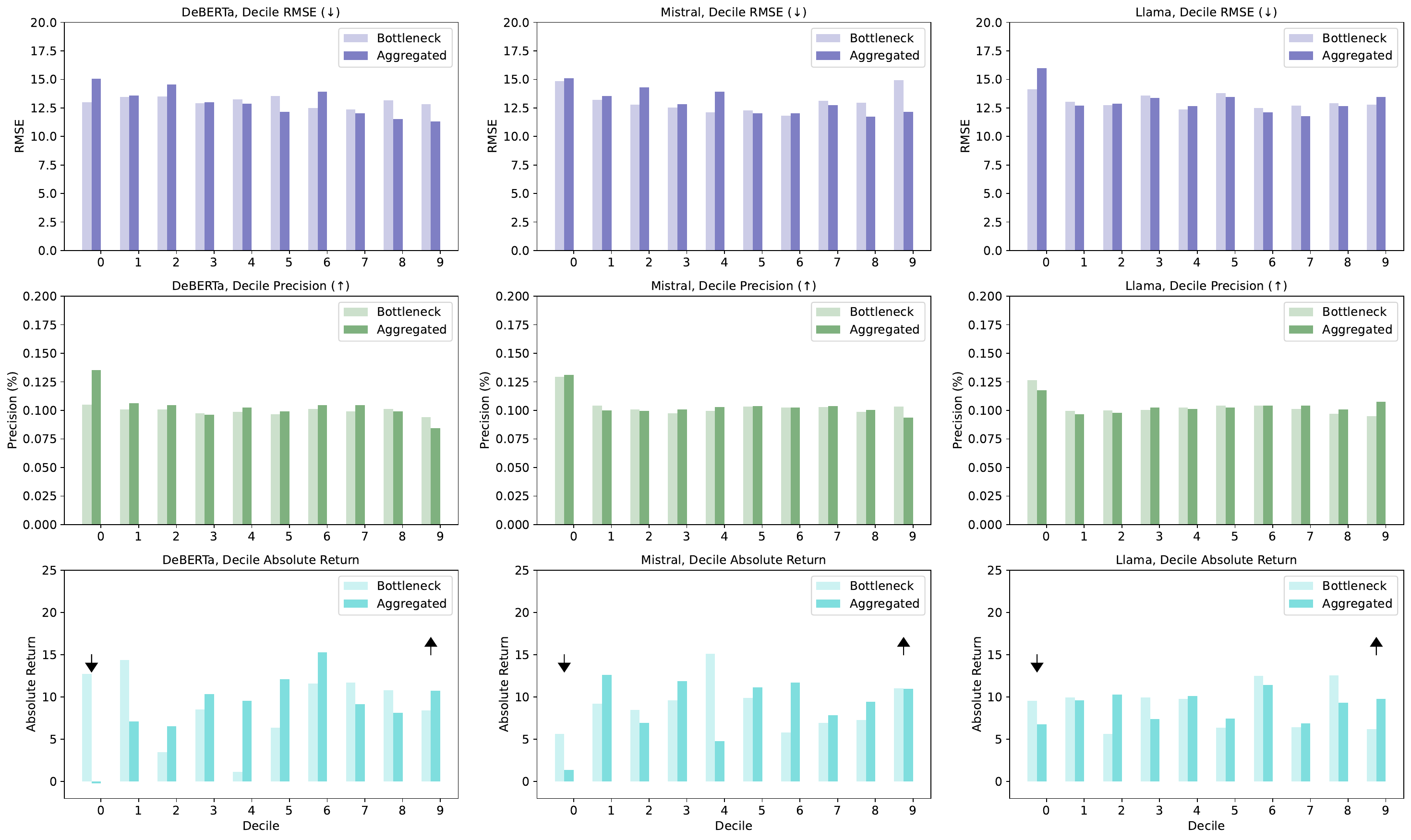}
    \end{subfigure}%
    \caption{
    Decile Performance of Bottleneck and Aggregated Representations in the Emerging Markets Universe (best viewed in color).
    Top Row: Decile RMSE.
    Middle Row: Decile Precision.
    Bottom Row: Decile Return.
    The up (or down) arrow indicates the higher (or lower) values are desirable.
    }
    \label{fig:decile_em}
\end{figure}

\begin{table*}[!htbp]
  \centering
  \caption{
  Statistics of Portfolios in the Emerging Markets Universe. 
  The Universe Equally-Weighted represents the universe performance reported under the Long-only Portfolio column.
  }
  \resizebox{1.\textwidth}{!}{
  \begin{tabular}{|c|c|c|c|c|}
    \hline
    \multirow{2}{*}{} & \multicolumn{2}{|c|}{Long-only Portfolio} & \multicolumn{2}{|c|}{Long-short Portfolio} \\
    \cline{2-5}
    & Ann. Return \% ($\uparrow$) & Sharpe Ratio ($\uparrow$) & Ann. Return \% ($\uparrow$) & Sharpe Ratio ($\uparrow$) \\
    \hline
    Universe Equally-Weighted      & 3.91 & 0.32 & $-$   & $-$ \\
    Sentiment\_FinVader            & 6.18 & 0.43 & -0.08 & 0.04 \\
    Sentiment\_FinBert             & 9.76 & 0.70 &  1.69 & 0.21 \\
    \hline
    DeBERTa\_Bottleneck  &  7.32 & 0.50 & -5.00 & -0.36 \\
    DeBERTa\_Aggregated  &  9.88 & 0.64 & 10.96 & 0.97 \\
    Mistral\_Bottleneck  & 10.12 & 0.63 & 4.94  & 0.47 \\
    Mistral\_Aggregated  & 10.11 & 0.64 & 9.16  & 0.68 \\
    Llama\_Bottleneck    &  4.94 & 0.36 & -3.99 & -0.28 \\
    Llama\_Aggregated    &  8.82 & 0.58 & 1.83  & 0.19  \\
    \hline
  \end{tabular}
  }
  \label{tab:portfolio_em}
\end{table*}

\begin{figure}[!htbp]
  \begin{center}
    \includegraphics[width=0.85\textwidth]{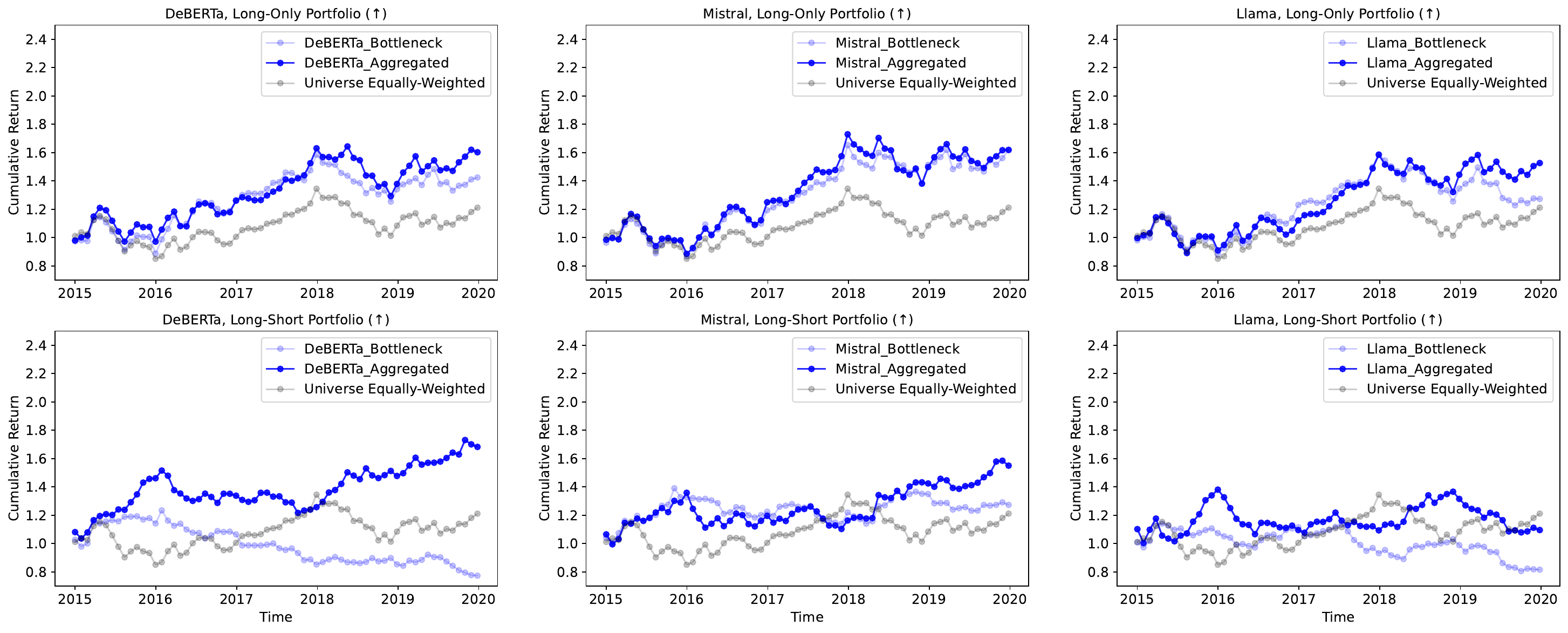}
  \end{center}
  \caption{
  Cumulative Return Charts of the Portfolios based on Bottleneck and Aggregated Representation Models in the Emerging Markets Universe (best viewed in color).
  Top Row: Long-only Portfolios. 
  Bottom Row: Long-short Portfolios.
  }
  \label{fig:portfolio_em}
\end{figure}

\clearpage
\begin{figure}[t]
  \begin{center}
    \includegraphics[width=0.85\textwidth]{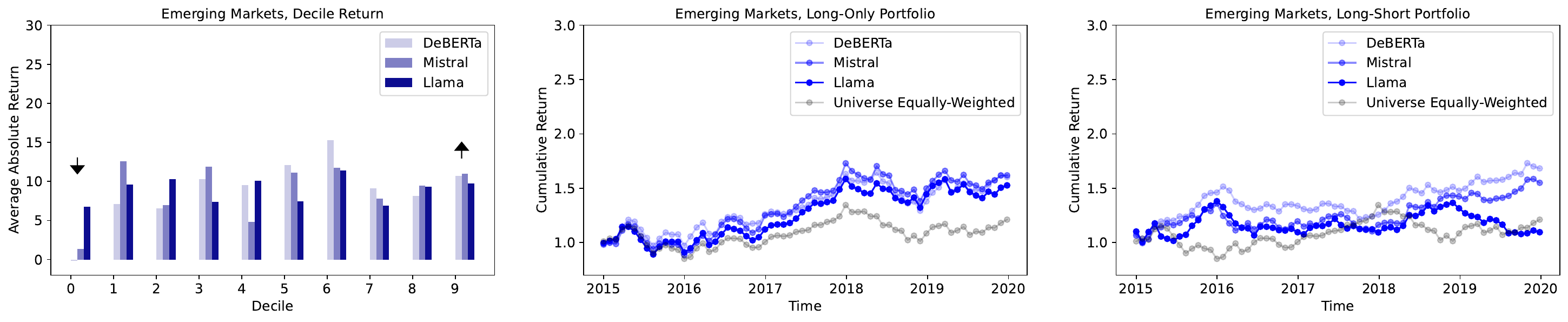}
  \end{center}
  \caption{
  Comparison of Encoder-only and Decoder-only LLMs with the Suited Representations in the Emerging Markets Universe (best viewed in color).
  }
  \label{fig:compare_enc_dec_em}
\end{figure}

\begin{figure}[htbp]
  \begin{center}
    \includegraphics[width=0.85\textwidth]{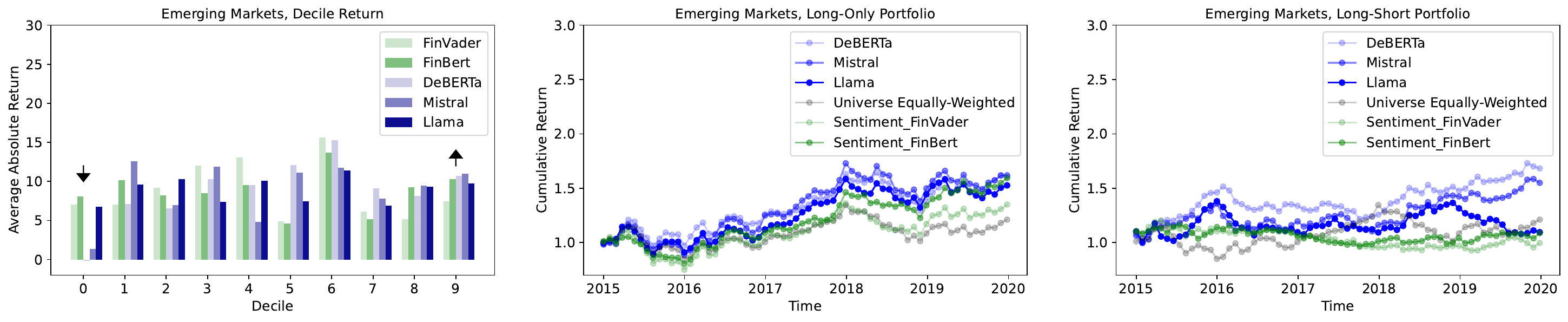}
  \end{center}
  \caption{
  Comparison with Sentiment-based Portfolios in the Emerging Markets Universe (best viewed in color).
  }
  \label{fig:compare_senti_em}
\end{figure}

%% file: main.bbl
\begin{thebibliography}{10}

\bibitem{allen2019daily}
David~E Allen, Michael McAleer, and Abhay~K Singh.
\newblock Daily market news sentiment and stock prices.
\newblock {\em Applied Economics}, 51(30):3212--3235, 2019.

\bibitem{ang2014asset}
Andrew Ang.
\newblock {\em Asset management: A systematic approach to factor investing}.
\newblock Oxford University Press, 2014.

\bibitem{araci2019finbert}
Dogu Araci.
\newblock Finbert: Financial sentiment analysis with pre-trained language models.
\newblock {\em arXiv preprint arXiv:1908.10063}, 2019.

\bibitem{barbero2024transformers}
Federico Barbero, Andrea Banino, Steven Kapturowski, Dharshan Kumaran, Jo{\~a}o~GM Ara{\'u}jo, Alex Vitvitskyi, Razvan Pascanu, and Petar Veli{\v{c}}kovi{\'c}.
\newblock Transformers need glasses! information over-squashing in language tasks.
\newblock {\em arXiv preprint arXiv:2406.04267}, 2024.

\bibitem{behnamghader2024llm2vec}
Parishad BehnamGhader, Vaibhav Adlakha, Marius Mosbach, Dzmitry Bahdanau, Nicolas Chapados, and Siva Reddy.
\newblock Llm2vec: Large language models are secretly powerful text encoders.
\newblock {\em arXiv preprint arXiv:2404.05961}, 2024.

\bibitem{brown2020language}
Tom Brown, Benjamin Mann, Nick Ryder, Melanie Subbiah, Jared~D Kaplan, Prafulla Dhariwal, Arvind Neelakantan, Pranav Shyam, Girish Sastry, Amanda Askell, et~al.
\newblock Language models are few-shot learners.
\newblock {\em Advances in neural information processing systems}, 33:1877--1901, 2020.

\bibitem{chen2019group}
Deli Chen, Keiko Harimoto, Ruihan Bao, Qi~Su, Xu~Sun, et~al.
\newblock Group, extract and aggregate: Summarizing a large amount of finance news for forex movement prediction.
\newblock {\em arXiv preprint arXiv:1910.05032}, 2019.

\bibitem{chung2024scaling}
Hyung~Won Chung, Le~Hou, Shayne Longpre, Barret Zoph, Yi~Tay, William Fedus, Yunxuan Li, Xuezhi Wang, Mostafa Dehghani, Siddhartha Brahma, et~al.
\newblock Scaling instruction-finetuned language models.
\newblock {\em Journal of Machine Learning Research}, 25(70):1--53, 2024.

\bibitem{devlin2019bert}
Jacob Devlin, Ming-Wei Chang, Kenton Lee, and Kristina Toutanova.
\newblock Bert: Pre-training of deep bidirectional transformers for language understanding.
\newblock In {\em Proceedings of the 2019 Conference of the North American Chapter of the Association for Computational Linguistics: Human Language Technologies, Volume 1 (Long and Short Papers)}, pages 4171--4186, 2019.

\bibitem{ding2023parameter}
Ning Ding, Yujia Qin, Guang Yang, Fuchao Wei, Zonghan Yang, Yusheng Su, Shengding Hu, Yulin Chen, Chi-Min Chan, Weize Chen, et~al.
\newblock Parameter-efficient fine-tuning of large-scale pre-trained language models.
\newblock {\em Nature Machine Intelligence}, 5(3):220--235, 2023.

\bibitem{fama1996multifactor}
Eugene~F Fama and Kenneth~R French.
\newblock Multifactor explanations of asset pricing anomalies.
\newblock {\em The journal of finance}, 51(1):55--84, 1996.

\bibitem{gunel2020supervised}
Beliz Gunel, Jingfei Du, Alexis Conneau, and Ves Stoyanov.
\newblock Supervised contrastive learning for pre-trained language model fine-tuning.
\newblock {\em arXiv preprint arXiv:2011.01403}, 2020.

\bibitem{guo2020esg2risk}
Tian Guo, Nicolas Jamet, Valentin Betrix, Louis-Alexandre Piquet, and Emmanuel Hauptmann.
\newblock Esg2risk: A deep learning framework from esg news to stock volatility prediction.
\newblock {\em arXiv preprint arXiv:2005.02527}, 2020.

\bibitem{he2021debertav3}
Pengcheng He, Jianfeng Gao, and Weizhu Chen.
\newblock Debertav3: Improving deberta using electra-style pre-training with gradient-disentangled embedding sharing.
\newblock {\em arXiv preprint arXiv:2111.09543}, 2021.

\bibitem{he2020deberta}
Pengcheng He, Xiaodong Liu, Jianfeng Gao, and Weizhu Chen.
\newblock Deberta: Decoding-enhanced bert with disentangled attention.
\newblock {\em arXiv preprint arXiv:2006.03654}, 2020.

\bibitem{hu2021lora}
Edward~J Hu, Yelong Shen, Phillip Wallis, Zeyuan Allen-Zhu, Yuanzhi Li, Shean Wang, Lu~Wang, and Weizhu Chen.
\newblock Lora: Low-rank adaptation of large language models.
\newblock {\em arXiv preprint arXiv:2106.09685}, 2021.

\bibitem{hu2018listening}
Ziniu Hu, Weiqing Liu, Jiang Bian, Xuanzhe Liu, and Tie-Yan Liu.
\newblock Listening to chaotic whispers: A deep learning framework for news-oriented stock trend prediction.
\newblock In {\em Proceedings of the eleventh ACM international conference on web search and data mining}, pages 261--269, 2018.

\bibitem{hutto2014vader}
Clayton Hutto and Eric Gilbert.
\newblock Vader: A parsimonious rule-based model for sentiment analysis of social media text.
\newblock In {\em Proceedings of the international AAAI conference on web and social media}, volume~8, pages 216--225, 2014.

\bibitem{jiang2023mistral}
Albert~Q Jiang, Alexandre Sablayrolles, Arthur Mensch, Chris Bamford, Devendra~Singh Chaplot, Diego de~las Casas, Florian Bressand, Gianna Lengyel, Guillaume Lample, Lucile Saulnier, et~al.
\newblock Mistral 7b.
\newblock {\em arXiv preprint arXiv:2310.06825}, 2023.

\bibitem{kim2024financial}
Alex Kim, Maximilian Muhn, and Valeri~V Nikolaev.
\newblock Financial statement analysis with large language models.
\newblock {\em Chicago Booth Research Paper Forthcoming, Fama-Miller Working Paper}, 2024.

\bibitem{kojima2022large}
Takeshi Kojima, Shixiang~Shane Gu, Machel Reid, Yutaka Matsuo, and Yusuke Iwasawa.
\newblock Large language models are zero-shot reasoners.
\newblock {\em Advances in neural information processing systems}, 35:22199--22213, 2022.

\bibitem{FinVADER}
Petr Korab.
\newblock Finvader: Financial sentiment analysis.
\newblock \url{https://github.com/PetrKorab/FinVADER}, 2023.

\bibitem{lan2019albert}
Zhenzhong Lan, Mingda Chen, Sebastian Goodman, Kevin Gimpel, Piyush Sharma, and Radu Soricut.
\newblock Albert: A lite bert for self-supervised learning of language representations.
\newblock In {\em International Conference on Learning Representations}, 2019.

\bibitem{lee2024nv}
Chankyu Lee, Rajarshi Roy, Mengyao Xu, Jonathan Raiman, Mohammad Shoeybi, Bryan Catanzaro, and Wei Ping.
\newblock Nv-embed: Improved techniques for training llms as generalist embedding models.
\newblock {\em arXiv preprint arXiv:2405.17428}, 2024.

\bibitem{lewis2019bart}
Mike Lewis, Yinhan Liu, Naman Goyal, Marjan Ghazvininejad, Abdelrahman Mohamed, Omer Levy, Ves Stoyanov, and Luke Zettlemoyer.
\newblock Bart: Denoising sequence-to-sequence pre-training for natural language generation, translation, and comprehension.
\newblock {\em arXiv preprint arXiv:1910.13461}, 2019.

\bibitem{li2023large}
Yinheng Li, Shaofei Wang, Han Ding, and Hang Chen.
\newblock Large language models in finance: A survey.
\newblock In {\em Proceedings of the fourth ACM international conference on AI in finance}, pages 374--382, 2023.

\bibitem{liu2018hierarchical}
Qikai Liu, Xiang Cheng, Sen Su, and Shuguang Zhu.
\newblock Hierarchical complementary attention network for predicting stock price movements with news.
\newblock In {\em Proceedings of the 27th ACM International Conference on Information and Knowledge Management}, pages 1603--1606, 2018.

\bibitem{liu2024dora}
Shih-Yang Liu, Chien-Yi Wang, Hongxu Yin, Pavlo Molchanov, Yu-Chiang~Frank Wang, Kwang-Ting Cheng, and Min-Hung Chen.
\newblock Dora: Weight-decomposed low-rank adaptation.
\newblock {\em arXiv preprint arXiv:2402.09353}, 2024.

\bibitem{liu2021finbert}
Zhuang Liu, Degen Huang, Kaiyu Huang, Zhuang Li, and Jun Zhao.
\newblock Finbert: A pre-trained financial language representation model for financial text mining.
\newblock In {\em Proceedings of the twenty-ninth international conference on international joint conferences on artificial intelligence}, pages 4513--4519, 2021.

\bibitem{qin2019you}
Yu~Qin and Yi~Yang.
\newblock What you say and how you say it matters: Predicting financial risk using verbal and vocal cues.
\newblock In {\em 57th Annual Meeting of the Association for Computational Linguistics (ACL 2019)}, page 390, 2019.

\bibitem{radfordimproving}
Alec Radford, Karthik Narasimhan, Tim Salimans, and Ilya Sutskever.
\newblock Improving language understanding by generative pre-training.
\newblock 2018.

\bibitem{radford2019language}
Alec Radford, Jeffrey Wu, Rewon Child, David Luan, Dario Amodei, Ilya Sutskever, et~al.
\newblock Language models are unsupervised multitask learners.
\newblock {\em OpenAI blog}, 1(8):9, 2019.

\bibitem{raffel2020exploring}
Colin Raffel, Noam Shazeer, Adam Roberts, Katherine Lee, Sharan Narang, Michael Matena, Yanqi Zhou, Wei Li, and Peter~J Liu.
\newblock Exploring the limits of transfer learning with a unified text-to-text transformer.
\newblock {\em Journal of machine learning research}, 21(140):1--67, 2020.

\bibitem{rasley2020deepspeed}
Jeff Rasley, Samyam Rajbhandari, Olatunji Ruwase, and Yuxiong He.
\newblock Deepspeed: System optimizations enable training deep learning models with over 100 billion parameters.
\newblock In {\em Proceedings of the 26th ACM SIGKDD International Conference on Knowledge Discovery \& Data Mining}, pages 3505--3506, 2020.

\bibitem{sawhney2020deep}
Ramit Sawhney, Shivam Agarwal, Arnav Wadhwa, and Rajiv Shah.
\newblock Deep attentive learning for stock movement prediction from social media text and company correlations.
\newblock In {\em Proceedings of the 2020 Conference on Empirical Methods in Natural Language Processing (EMNLP)}, pages 8415--8426, 2020.

\bibitem{shapiro2022measuring}
Adam~Hale Shapiro, Moritz Sudhof, and Daniel~J Wilson.
\newblock Measuring news sentiment.
\newblock {\em Journal of econometrics}, 228(2):221--243, 2022.

\bibitem{touvron2023llama}
Hugo Touvron, Louis Martin, Kevin Stone, Peter Albert, Amjad Almahairi, Yasmine Babaei, Nikolay Bashlykov, Soumya Batra, Prajjwal Bhargava, Shruti Bhosale, et~al.
\newblock Llama 2: Open foundation and fine-tuned chat models.
\newblock {\em arXiv preprint arXiv:2307.09288}, 2023.

\bibitem{wang2023simlm}
Liang Wang, Nan Yang, Xiaolong Huang, Binxing Jiao, Linjun Yang, Daxin Jiang, Rangan Majumder, and Furu Wei.
\newblock Simlm: Pre-training with representation bottleneck for dense passage retrieval.
\newblock In {\em The 61st Annual Meeting Of The Association For Computational Linguistics}, 2023.

\bibitem{wang2023improving}
Liang Wang, Nan Yang, Xiaolong Huang, Linjun Yang, Rangan Majumder, and Furu Wei.
\newblock Improving text embeddings with large language models.
\newblock {\em arXiv preprint arXiv:2401.00368}, 2023.

\bibitem{wang2024llmfactor}
Meiyun Wang, Kiyoshi Izumi, and Hiroki Sakaji.
\newblock Llmfactor: Extracting profitable factors through prompts for explainable stock movement prediction.
\newblock {\em arXiv preprint arXiv:2406.10811}, 2024.

\bibitem{wang2019ean}
Yaowei Wang, Qing Li, Zhexue Huang, and Junjie Li.
\newblock Ean: Event attention network for stock price trend prediction based on sentimental embedding.
\newblock In {\em Proceedings of the 10th ACM Conference on Web Science}, pages 311--320, 2019.

\bibitem{wei2021finetuned}
Jason Wei, Maarten Bosma, Vincent~Y Zhao, Kelvin Guu, Adams~Wei Yu, Brian Lester, Nan Du, Andrew~M Dai, and Quoc~V Le.
\newblock Finetuned language models are zero-shot learners.
\newblock {\em arXiv preprint arXiv:2109.01652}, 2021.

\bibitem{wei2022chain}
Jason Wei, Xuezhi Wang, Dale Schuurmans, Maarten Bosma, Fei Xia, Ed~Chi, Quoc~V Le, Denny Zhou, et~al.
\newblock Chain-of-thought prompting elicits reasoning in large language models.
\newblock {\em Advances in neural information processing systems}, 35:24824--24837, 2022.

\bibitem{weng2018predicting}
Bin Weng, Lin Lu, Xing Wang, Fadel~M Megahed, and Waldyn Martinez.
\newblock Predicting short-term stock prices using ensemble methods and online data sources.
\newblock {\em Expert Systems with Applications}, 112:258--273, 2018.

\bibitem{xu2018stock}
Yumo Xu and Shay~B Cohen.
\newblock Stock movement prediction from tweets and historical prices.
\newblock In {\em Proceedings of the 56th Annual Meeting of the Association for Computational Linguistics (Volume 1: Long Papers)}, pages 1970--1979, 2018.

\bibitem{yang2019leveraging}
Linyi Yang, Ruihai Dong, Tin Lok~James Ng, and Yang Xu.
\newblock Leveraging bert to improve the fears index for stock forecasting.
\newblock In {\em Proceedings of the First Workshop on Financial Technology and Natural Language Processing}, pages 54--60, 2019.

\bibitem{zhao2023survey}
Wayne~Xin Zhao, Kun Zhou, Junyi Li, Tianyi Tang, Xiaolei Wang, Yupeng Hou, Yingqian Min, Beichen Zhang, Junjie Zhang, Zican Dong, et~al.
\newblock A survey of large language models.
\newblock {\em arXiv preprint arXiv:2303.18223}, 2023.

\end{thebibliography}
